\documentclass[twocolumn,showpacs,preprintnumbers,amsmath,amssymb]{revtex4}
\usepackage{graphicx}
\usepackage{dcolumn}
\usepackage{bm}

\begin{document}


\title{Regge-trajectory Analysis of $D^{\star}_{SJ}(2317)^\pm$, $D_{SJ}(2460)^\pm$ and
$D_{SJ}(2632)^+$ Mesons}
\author{Ailin Zhang}
\affiliation{ Department of Physics, Shanghai University, Shanghai,
200444, China}


\begin{abstract}
Status of investigations of the new observed charmed strange mesons
$D^{\star}_{SJ}(2317)^\pm$, $D_{SJ}(2460)^\pm$ and $D_{SJ}(2632)^+$
is simply reviewed. A systemic classification to these states with
Regge trajectories(RTs) was made. We found that
$D^{\star}_{SJ}(2317)^\pm$ and $D_{SJ}(2460)^\pm$ are reasonable to
be arranged as $(0^+,1^+)$ states, but $D_{SJ}(2632)^+$ seems not
possible to be an orbital excited tensor particle. As a byproduct,
the non-strange charmed mesons including $D^\prime_1(2427)$ and
$D^\star(2637)^+$ were analyzed also.
\end{abstract}

\maketitle

The problem of Quantum Chromodynamics(QCD) spectrum is a central
issue in nonperturbative QCD and is connected to problems of
confinement and mass generation, the flavor dependence of hadron
spectrum and its connection to the type of potentials are still not
clear. Charmed strange meson is an important system to study hadron
spectrum for its internal heavy-light quark(anti-quark) components.
There were limited experimental data for these mesons before, but
the situation changes a lot since last year for the observation of
several new states.

$D^{\star}_{SJ}(2317)^\pm$ was first observed in $D^+_S\pi$ by
BaBar\cite{babar}, then confirmed by CLEO\cite{cleo},
BELLE\cite{belle} and FOCUS\cite{focus}. This state has mass
$2317.4\pm 0.9$ MeV from PDG\cite{pdg}, about $40$ MeV below $DK$
threshold, and has full width $\Gamma<4.6$ MeV at $90\%$ confidence
level.

$D_{SJ}(2460)^\pm$ was first reported by CLEO\cite{cleo} in
$D^\star_S\pi^0$ final states, and later observed by
BELLE\cite{belle1} and BaBar\cite{babar1}. This state has mass
$2459.3\pm 1.3$ MeV\cite{pdg}, about $50$ MeV below $D^\star K$
threshold, and has full width $\Gamma<5.5$ MeV at $90\%$ CL.

Very recently, a new surprisingly narrow charmed strange meson,
$D_{SJ}(2632)^+$, was reported by SELEX\cite{selex} in $D^+_S\eta$
and $D^0K^+$ decay channels. The reported mass is $2632.6\pm 1.6$
MeV, about $274$ MeV and $116$ MeV above $D^0K^+$ and $D_S\eta$
threshold, respectively. It has width $\Gamma<17$ MeV at $90\%$ CL.
This state has an exotic relative branching ratio
$\Gamma(D^0K^+)/\Gamma(D^+_S\eta)=0.16\pm 0.06$.

Spectrum of heavy-light system has been studied with many
theoretical methods. In a unified quark model, the charmed strange
$c\bar s$ mesons have been computed by
Godfrey-Isgur-Kokoski\cite{gik}. The predicted masses are higher
than the observed experimental data. In a relativistic quark
model, the orbitally and the radially excited D and B mesons was
calculated\cite{ebert}. The predicted masses are lower than the
observed experimental data.

In addition to computation of the spectrum of S-wave and P-wave
heavy light mesons with NRQCD on lattice in quenched approximation,
computation of the spectrum of radially and orbitally excited states
has also been made\cite{lewis}. Lattice predicted masses in these
calculations are higher than experimental results.

In Heavy Quark Effective Theory (HQET)\cite{hqet}, the spin of
heavy quark decouples from the rest of the system and the meson
properties are determined alone by light degrees of freedom. It is
then possible to classify the heavy light hadrons with the light
degrees of freedom of spin $j$ as a good quantum number.
Therefore, heavy light mesons can be collected in doublets with a
peculiar spin and parity $j^P$. Members of each doublet degenerate
in mass in leading order. Spin-parity of ground states in this
model is $j^P={1\over 2}^-$. This doublet comprises two states
with total spin-parity $J^P=(0^-,1^-)$, which correspond to
$^1S_0$ and $^3S_1$ states in normal quark models. The first
excited states involving a P-wave excitation have light degrees of
freedom $j^P={1\over 2}^+$ or ${3\over 2}^+$. The two doublets
have $J^P=(0^+,1^+)$ and $J^P=(1^+,2^+)$, respectively.

When the spontaneous breaking of chiral symmetry was incorporated
in, heavy light system could be studied in a chiral quark
model\cite{chiral}. In this model, the mesons are predicted to
appear in parity-doubled bound states, which transform as linear
representations of the light quark chiral symmetry. Parity doublet
has the same mass splittings. For low lying mesons, the chiral
partner of ground state $(0^-,1^-)$  is excited state $(0^+,1^+)$ .
The spectrum of corresponding excited D mesons have been calculated
by W. Bardeen {\it et al}\cite{bardeen} recently. Their results are
in good agreement with experimental data.

Calculations with other methods such as QCD string, unitarized
meson model, MIT bag model and QCD sum rules\cite{other} will not
be introduced here.

Based on computations of the spectra and analyses to their decays,
enormous discussions about the nature of these states have been
triggered. $D^{\star}_{SJ}(2317)^\pm$ was explained as $DK$ meson
molecule and $D\pi$ atom\cite{molecule}, four quark
state\cite{four}, P-wave $^3P_0$ $c\bar s$
mesons\cite{bardeen,other,cs}, baryonium\cite{datta} and mixed
state\cite{mixed}. $D_{SJ}(2460)^\pm$ has a similar explanation
except for the P-wave $^1P_1$ $c\bar s$ explanation.
$D_{SJ}(2632)^+$ was suggested to be a four quark
state\cite{four1}, tetraquarks\cite{tetra} and the first radial
excitation state of $D^\star_S(2112)^\pm$\cite{radial}. A
systematic review to this excited subject could be found in
\cite{review}.

So far, all the calculations of hadron spectrum have relied on
some models. In this paper, we will make a phenomenological
analysis to these excited states by means of approximate linear
structure of the Regge trajectories and will make a systemic
classification to them. In fact, if the new data about these
resonances has been confirmed, it is possible to study the
properties of their Regge trajectories.

Several decades ago, it was known from meson phenomenology that
the square of the hadron masses depend approximately linearly on
the spin of the hadrons, which resulted in Regge trajectories
theory. A Regge trajectory is a line in a
Chew-Frautschi\cite{chew} plot representing the spin of the
lightest particles of that spin versus their mass square, t:
\begin{eqnarray}
\alpha(t)=\alpha(0)+\alpha^\prime t
\end{eqnarray}
where intercept $\alpha(0)$ and slope $\alpha^\prime$ depend weekly
on the flavor content of the state lying on corresponding
trajectory. A Regge trajectory is approximately linear, while
different trajectories are approximately parallel.

Based on much trial and experimentation, the flavor dependence of
$q\bar q$ mesons was assumed to be on quark masses combination
$m_1+m_2$. A global description to Regge trajectories for all
flavors was constructed\cite{fili}
\begin{eqnarray}\label{rt}
\alpha(m_1+m_2,t)=\alpha_I(m_1+m_2,0)+\alpha^\prime(m_1+m_2) t,
\end{eqnarray}
where the subscript I refers to the leading trajectory.

When the mesons for which the lowest physical state is at $J=1$
are concerned,
\begin{eqnarray}\label{slope}
\alpha_I(m_1+m_2,0)&=&0.57-{(m_1+m_2)\over GeV},\\\nonumber
\alpha^\prime(m_1+m_2)&=&{0.9~GeV^{-2}\over [1+0.22({m_1+m_2\over
GeV})^{3/2}]}.
\end{eqnarray}
For light quark mesons, $ \alpha^\prime\approx 0.9~GeV^{-2}$. For
leading trajectories whose ground states begin at $J=0$, they have
an intercept approximately $0.5$ MeV lower and follow a similar
pattern.

For radial excited light $q\bar q$ mesons, trajectories on
$(n,M^2)$-plots are obtained by\cite{ani}
\begin{eqnarray}\label{nm}
M^2=M^2_0+(n-1)\mu^2,
\end{eqnarray}
where $M_0$ is the mass of basic meson, n is the radial quantum
number, and $\mu^2$ (approximately the same for all trajectories)
is the slope parameter of the trajectory.

Properties of Regge trajectories of baryons\cite{baryon},
glueballs\cite{glueball} and hybrids\cite{hybrid} have also been
studied in many references.

Eq. [\ref{rt}] was constructed from a comprehensive
phenomenological analysis of available experimental data for
mesonic resonances of light, medium and heavy flavors. It has been
supplemented by results from various phenomenological models.

As well known, a Regge trajectorie may deviate from straight line,
and different trajectories may deviate from
parallelism\cite{nonlinear}. The exact deviation depends on peculiar
family of mesons, baryons, glueballs, hybrids and energy region. In
fact, the non-linearity and non-parallelism of RT depends on
intrinsic quark-gluon dynamics including flavor and J dependence
though the exact intrinsic dynamics is unknown. More detailed
studies of Regge trajectories have been made in many more
fundamental theories\cite{string}.

However, for mesons with small J, spin-orbit contribution is not
significant, once the flavor dependence is the same, intrinsic
dynamics is similar. Therefore the linearity and the parallelism of
Regge trajectories are kept well. In the mean time, deviation from
exchange degeneracy could not be large.

Based on these analyses and Eq. [\ref{rt}][\ref{slope}], the
linearity, the parallelism and the masses combination $m_1+m_2$
dependence(flavor dependence) of Regge trajectories for heavy light
mesons with small J are assumed in this paper. By means of these
assumptions, we start our analysis to the spectrum of mesons.

In quark model, $q\bar q$ mesons could be marked by their quantum
numbers, $In^{2S+1}L_J$. From PDG\cite{pdg}, we get
Table~\ref{table-1} for charmed strange mesons. In this table,
entries in the first volume are observed mesons, entries in the last
volume are information from PDG, entries under $J^P$, $n^{2S+1}L_J$
and $j^P$(light degrees of freedom) for those unconfirmed mesons are
favored assignment by theoretical analyses.
\begin{table}
\begin{tabular}{lllllll}
 States & $J^P$ &  $n^{2S+1}L_J$ & $j^p$
& PDG note\\
\hline\hline $D_S(1969)^\pm$ & $0^-$ & $1^1S_0$ & ${1\over 2}^-$
& \\
$D^\star_S(2112)^\pm$ & $1^-$ & $1^3S_1$ & ${1\over 2}^-$ & $J^P=?^?$ consistent with $1^-$ \\
\hline\hline $D^\star_{SJ}(2317)^\pm$& $0^+$ & $1^3P_0$ & ${1\over
2}^+$
& J, P need confirmation \\
$D_{SJ}(2460)^\pm$& $1^+$ & $1^1P_1$ &
${1\over 2}^+$         \\
\hline\hline $D_{S1}(2536)^\pm$& $1^+$ & $1^3P_1$ & ${3\over
2}^+$ & J, P need confirmation \\
$D_{S2}(2573)^\pm$& $2^+$ & $1^3P_2$ & ${3\over 2}^+$ &
$J^P=?^?$ consistent with $2^+$ \\
\hline\hline $D_{SJ}(2632)^+$& $1^-$ & $2^3S_1$ & ${1\over
2}^-$ & $J^P=?^?$\\
\hline\hline
\end{tabular}
\caption{Spectrum of Charmed and Strange Mesons.} \label{table-1}
\end{table}

In chiral quark model, the new observed $D^\star_{SJ}(2317)^\pm$,
$D_{SJ}(2460)^\pm$ are suggested to be $(0^+,1^+)$ states, the
chiral doubler of $(0^-,1^-)$ states: $D_S(1969)^\pm$ and
$D^\star_S(2112)^\pm$. They have similar splitting $\approx 348$
MeV:
\begin{eqnarray}
D_{SJ}(2460)^\pm -D^\star_S(2112)^\pm \approx
D^\star_{SJ}(2317)^\pm -D_S(1969)^\pm
\end{eqnarray}

\begin{figure}
\includegraphics{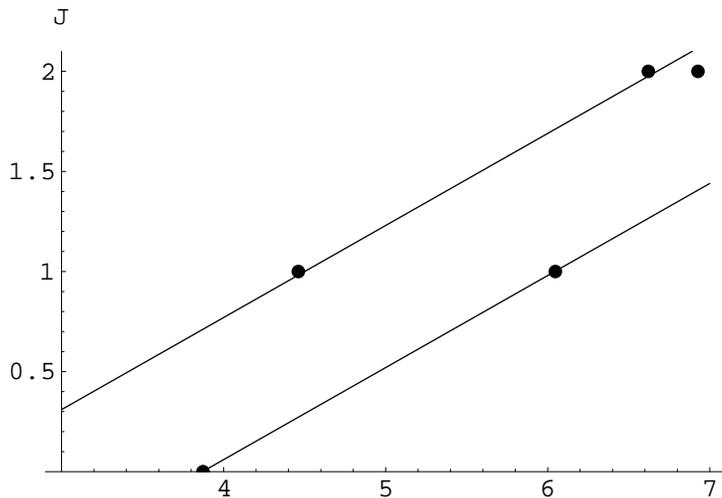}
\caption{\label{rj1} Chew-Frautschi plots (t,J) for
$D_S(1969)^\pm(0^-$), $D^\star_S(2112)^\pm(1^-$),
$D_{SJ}(2460)^\pm$ ($1^+$) and $D_{S2}(2573)^\pm$ ($2^+$), where
the $D_{SJ}(2632)^+$ lies outside the straight line.}
\end{figure}

Let us check this assignment with Regge trajectories. As well known,
when the deviation from exchange degeneracy is not large, the
$D^\star_S(2112)^\pm$ ($1^-$) RT and the $D_{S2}(2573)^\pm$ ($2^+$)
RT is almost the same and they determine a unique trajectory with
slope
\begin{eqnarray}
\alpha^\prime(m_c+m_s)={1\over 2.573^2-2.112^2}GeV^{-2}\approx
0.464~GeV^{-2}.
\end{eqnarray}
$D_S(1969)^\pm$ ($0^-$) and $D_{SJ}(2460)^\pm (1^+$) determine
another trajectory with slope
\begin{eqnarray}
\alpha^\prime(m_c+m_s)={1\over 2.459^2-1.968^2}GeV^{-2}\approx
0.460~GeV^{-2}.
\end{eqnarray}
The slopes of two trajectories are approximately the same and two
trajectories are parallel(a natural conclusion of Eq.
[\ref{slope}]). Our simple analysis supports the assignment for
mesons: $D_S(1969)^\pm$ ($0^-$), ~$D^\star_S(2112)^\pm$ ($1^-$),
~$D_{SJ}(2460)^\pm$ ($1^+$), ~$D_{S2}(2573)^\pm$ ($2^+$).
Correspondingly, the Chew-Frautschi plots were drawn in
Fig.~\ref{rj1}.

It is found that there exists no phenomenon called as spin-orbit
inversion\cite{ebert,isgur}, which may have relation with the
dynamics spin-dependence of the confinement.

When the new reported $D_{SJ}(2632)^+$ is assigned as the orbitally
excited $2^+~^3P_2$ state, $D^\star_S(2112)^\pm$ ($1^-$) and
$D_{SJ}(2632)^+$ ($2^+$) make a trajectory with slope
\begin{eqnarray}
\alpha^\prime(m_c+m_s)={1\over 2.632^2-2.112^2}GeV^{-2}\approx
0.405~GeV^{-2}.
\end{eqnarray}
The slope is much smaller than previous $0.460~GeV^{-2}$. Obviously,
if this assignment were right, deviation from parallelism of the two
trajectories with the same flavor would be large. There is no masses
dependence as Eq. [\ref{slope}] either. Therefore, once states
$D_S(1969)^\pm$ ($0^-$), ~$D^\star_S(2112)^\pm$ ($1^-$) and
$D_{SJ}(2460)^\pm$ ($1^+$) are confirmed by experiments, the
assignment of $D_{SJ}(2632)^+$ as a $2^+~^3P_2$ tensor resonance
seems impossible.

Now let us pay attention to the non-strange charmed mesons.
Information of the observed non-strange charmed states are
collected in Table~\ref{table-2}.
\begin{table}
\begin{tabular}{lllllll}
 States & $J^P$ &  $n^{2S+1}L_J$ & $j^p$
& PDG note\\
\hline\hline $D(1869)^\pm$ & $0^-$ & $1^1S_0$ & ${1\over 2}^-$
& \\
$D^\star(2010)^\pm$ & $1^-$ & $1^3S_1$ & ${1\over 2}^-$ & J, P need confirmation \\
\hline\hline $D^\star_0(2308)^\pm$& $0^+$ & $1^3P_0$ & ${1\over
2}^+$
& ? \\
$D^\prime_1(2427)$& $1^+$ & $1^1P_1$ &
${1\over 2}^+$ & ?        \\
\hline\hline $D_1(2420)^0$& $1^+$ & $1^3P_1$ & ${3\over
2}^+$ & J, P need confirmation \\
$D_2(2460)^\pm$& $2^+$ & $1^3P_2$ & ${3\over 2}^+$ &
$J^P=2^+$ strongly favored \\
\hline\hline $D^\star(2637)^+$& $1^-$ & $2^3S_1$ & ${1\over
2}^-$ & $J^P=?^?$\\
\hline\hline
\end{tabular}
\caption{Spectrum of Non-strange Charmed Mesons.} \label{table-2}
\end{table}
The $D^\star(2010)^\pm$ ($1^-)$ and $D_2(2460)^\pm$ ($2^+)$ make a
trajectory with slope
\begin{eqnarray}
\alpha^\prime(m_c+m_{u,d})={1\over 2.459^2-2.01^2}GeV^{-2}\approx
0.498~GeV^{-2}.
\end{eqnarray}
$\alpha^\prime(m_c+m_{u,d})$ is bigger than
$\alpha^\prime(m_c+m_s)$. It is obvious that the slopes of Regge
trajectories decrease with increasing quark mass. The obtained
result here about slopes supports the flavor dependence of Eq.
[\ref{slope}].

$D(1869)^\pm$ is the $0^-~^1S_0$ state, but the $1^+~^1P_1$ is
missing! Recently, the new observed $D^\star_0(2308)$ and
$D^\prime_1(2427)$\cite{belle2} were suggested as the $(0^+,1^+)$
chiral doubler of $(0^-,1^-)$ states: $D(1869)^\pm$ and
$D^\star(2010)^\pm$\cite{review}. If $D^\prime_1(2427)$ were the
missing $1^+~^1P_1$ state, then $D(1869)^\pm$ ($0^-$) and
$D^\prime_1(2427)$ ($1^+$) would make a trajectory with slope
\begin{eqnarray}
\alpha^\prime(m_c+m_{u,d})={1\over 2.427^2-1.869^2}GeV^{-2}\approx
0.417~GeV^{-2},
\end{eqnarray}
which is much smaller than previous $0.498~GeV^{-2}$. Obviously,
this assignment of $D^\prime_1(2427)$ is inconsistent with the
approximate linearity, the parallelism and the flavor dependence of
Regge trajectories. From the linearity, the parallelism and the
flavor dependence of Regge trajectories, the missing $1^+~^1P_1$
state should have mass $\approx 2350$ MeV.

\begin{figure}
\includegraphics{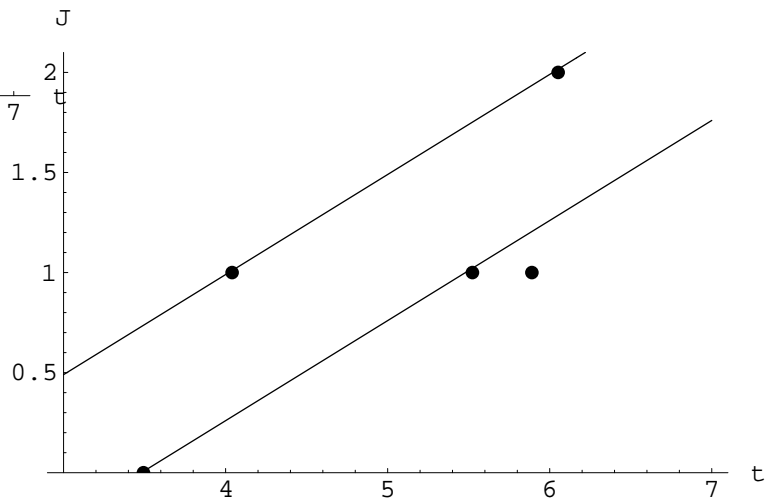}
\caption{\label{rj2} Chew-Frautschi plots (t,J) for $D(1869)^\pm$
($0^-)$, $D^\star(2010)^\pm(1^-)$, $D_2(2460)^\pm(2^+)$ and a
$1^+~^1P_1$ state $\approx 2350$ MeV, where the $D^\prime_1(2427)$
lies outside the straight line.}
\end{figure}

Similar to $D_{SJ}(2632)^+$, the recently observed
$D^\star(2637)^+$ by DELPHI in the $D^\star\pi\pi$
channel\cite{delphi} seems impossible to be assigned as a tensor
state. The Chew-Frautschi plots for these mesons were drawn in
Fig.~\ref{rj2}.

From the final states in its decay, $D_{SJ}(2632)^+$ must have
$J^P=0^+,~1^-,~2^+,\dots$. So this state has been
suggested\cite{radial} as the first radial excited state of
$D^\star_S(2112)^\pm (1^-)$. $D^\star(2637)^+$ was suggested as the
first radial excited states of $D^\star(2010)^\pm$
($1^-)$\cite{radial1}. If $D^\star(2637)^+$, $D_{SJ}(2632)^+$ are
really the first radial excited states of $D^\star(2010)^\pm$
($1^-)$, $D^\star_S(2112)^\pm (1^-)$, their spectra are exotic:
$1^-~^3S_1$ non-strange charmed meson lies below corresponding
charmed strange meson, but the first radial excited non-strange
charmed state lies above corresponding charmed strange meson.
Furthermore, their trajectories on $(n,M^2)$-plots are not
consistent with Eq. [\ref{nm}] for light mesons.

In conclusion, some interesting results on the charmed strange and
non-strange mesons have been obtained:

1, The slopes of the Regge trajectories decrease with increasing
quark mass, which is consistent with Eq.~[\ref{slope}].

2, The assignment of $D_{SJ}(2460)^\pm$ as $1^+~^1P_1$ state is
reasonable while the assignment of $D^\prime_1(2427)$ as
$1^+~^1P_1$ state seems impossible. The mass of the right
candidate of $1^+~^1P_1$ non-strange charmed state is predicted to
have mass $\approx 2350$ MeV.

3, The assignment of $D_{SJ}(2632)^+$ and $D^\star(2637)^+$ as the
$2^+~^3P_2$ state seems impossible.

4, If $D^\star(2637)^+$, $D_{SJ}(2632)^+$ are really the first
radial excited states, their spectra are exotic and their Regge
behavior is different from corresponding one for light mesons.

However, when we turn back to look at the entries in
Table~\ref{table-1} and \ref{table-2}, we find that we still have
little knowledge to heavy light charmed mesons. Quantum numbers of
some states are required to be measured, or to be confirmed. Some
predicted states should be searched for, and more decays modes
should be detected. We hope the investigation here will be useful to
further experiments.

The linearity, the parallelism and the flavor dependence of Regge
trajectories have been assumed in our analysis, these properties for
other mesons and possible deviations(and their origin) deserve more
study. If the approximate linearity, parallelism and the flavor
dependence of Regge trajectories of charmed mesons are confirmed
when more experimental data are accumulated, more hints about
mesons' intrinsic flavor dependence of their spectrum and about the
type of confinement potential for heavy light systems would be
discerned. Furthermore, reasonable conclusions from Regge
phenomenology are hoped to be incorporated into the study of quark
models.

Acknowledgment: This work is supported by National Natural Science
Foundation of China.

\end{document}